\documentclass[conference]{IEEEtran}
\IEEEoverridecommandlockouts
\usepackage{cite}
\usepackage{amsmath,amssymb,amsfonts}
\usepackage{algorithmic}
\usepackage{graphicx}
\usepackage{textcomp}
\usepackage{xcolor}
\usepackage[utf8]{inputenc}
\usepackage{amsmath}
\usepackage[english]{babel}
\usepackage{amsthm}
\usepackage{mathtools}
\usepackage{multirow}
\usepackage{mathtools}
\usepackage[ruled,norelsize]{algorithm2e}
\usepackage{bm} 
\usepackage{tikz}
\usepackage{subcaption}

\theoremstyle{definition}

\DeclareMathOperator\arctanh{arctanh}

\DeclarePairedDelimiter\floor{\lfloor}{\rfloor}

\def\BibTeX{{\rm B\kern-.05em{\sc i\kern-.025em b}\kern-.08em
    T\kern-.1667em\lower.7ex\hbox{E}\kern-.125emX}}

\makeatletter
\newcommand{\removelatexerror}{\let\@latex@error\@gobble}
\begin{document}
\title{Simple Gray Coding and LLR Calculation for MDS Modulation Systems\\
\thanks{The authors wish to acknowledge the support of the Bristol Innovation $\&$ Research Laboratory of Toshiba Research Europe Ltd. }
}

\author{\IEEEauthorblockN{Ferhat Yarkin and Justin P. Coon}
\IEEEauthorblockA{Department of Engineering Science, University of Oxford, Parks Road, Oxford, UK, OX1 3PJ \\
E-mail: \{ferhat.yarkin and justin.coon\}@eng.ox.ac.uk}
}

\maketitle

\begin{abstract}
Due to dependence between codeword elements, index modulation (IM) and related modulation techniques struggle to provide simple solutions for practical problems such as Gray coding between information bits and constellation points; and low-complexity log-likelihood ratio (LLR) calculations for channel-encoded information bits. In this paper, we show that a modulation technique based on a simple maximum distance separable (MDS) code, in other words, MDS modulation, can provide simple yet effective solutions to these problems, rendering the MDS techniques more beneficial in the presence of coding. We also compare the coded error performance of the MDS methods with that of the IM methods and demonstrate that MDS modulation outperforms IM.
\end{abstract}


\section{Introduction}

 Index modulation (IM) techniques that embed information into combinations/permutations  of codeword elements are considered to be promising modulation techniques due to their desirable properties in terms of error performance, low-complexity implementation, and compatibility with existing methods in space, time, and frequency domains \cite{Basar2013,Fan2015,Yarkin2021,Basar2015,Mao2017,Wen2017,Yarkin2020set}.

Despite the advantages of the IM techniques over conventional modulation techniques, it is not always easy to solve practical problems in these techniques. { For instance, the studies in the literature show that Gray mapping between index symbols and information bits is not straightforward. Although one can find low-complexity Gray coding algorithms for IM in the literature \cite{Xiao2020}, the IM systems using these algorithms still need to employ optimum maximum likelihood (ML) detection that exhibits excessive complexity when the codebook size is large. On the other hand, to calculate log-likelihood ratio (LLR) values related to information bits of the IM techniques, we need to consider all IM codewords due to the dependence between the codeword elements. The complexity of the LLR calculations will still be high when the codebook size is large since there is no direct way to calculate element-wise low-complexity LLR values for the IM systems. In this regard, the use of other mechanisms such as sphere decoding is necessary to make the calculations less complex \cite{Zeng2018}.} Such issues may limit the applications of the IM schemes for 5G and beyond 5G systems.  

On the other hand, modulation concepts based on a well-known maximum distance separable (MDS) code are capable of outperforming the IM concepts in terms of error performance while having even less implementation complexity \cite{Yarkin2020,Yarkin2021tcom}. These concepts use a simple MDS code over disjoint constellations to form codewords. Moreover, the MDS coding mechanism enables them to have more desirable properties than the IM techniques in error performance and low-complexity implementation. However, it is not evident whether such a mechanism can maintain its superiority against the IM approach when we have coding in the system. 

Against this background, in this paper, we study modulation concepts based on a well-known MDS code, i.e., MDS modulation, and show that these concepts can be more useful than the IM techniques in the presence of Gray Coding and channel encoding. { In this regard, we present element-wise low-complexity implementations are possible for Gray coding and LLR calculations of the MDS modulation techniques.} Then, we demonstrate by our numerical results that the MDS techniques can achieve better coded error performance. 

The rest of the paper is organized as follows. In Section \ref{sec:sec2}, we review the MDS modulation techniques. Simple Gray coding and LLR calculations are described for the MDS systems in Sections \ref{sec:sec3} and \ref{sec:sec4}, respectively. We present and compare numerical results in Section \ref{sec:sec5}. Then, the paper is concluded in \ref{sec:sec6}.

\section{MDS Modulation}\label{sec:sec2}

In this section, we review the structure of the MDS modulation techniques.

\subsection{Transmitter}

We use a simple MDS code along with $Q$ disjoint $M_1$-ary constellations to construct the $N$-dimensional complex constellation points, i.e., $\textbf{s}_1,\ldots, \textbf{s}_M$, where $N \ge 2$ and $M$ is the constellation size.\footnote{In this paper, we neglect the design of time-domain waveforms and focus on corresponding signal space representations where we have a signal constellation that is comprised of the vectors $\textbf{s}_1,\ldots, \textbf{s}_M$ as constellation points in an $N$-dimensional complex space. In this way, we aim to make the present study valid for any orthogonal transmission scheme in any domain.}   The MDS code forms the first $N-1$ elements of an $N$-tuple codeword, i.e., $I_{1}, I_{2} \ldots, I_{N-1}$, by using the integers, $1,\ldots,Q$ as symbols, i.e., $I_{\tau}\in \big\{1, \ldots, Q\big\}$ and $\tau\in \big\{1, \ldots, N-1\big\}$. The last element, $I_{N}$, is selected from the same integers by letting the code be those $N$-tuples summing to zero under modulo-$Q$ arithmetic, i.e., $(I_{1}+I_{2}+\ldots+I_{N}) \bmod Q=0 $ \cite{Singleton64}.  Therefore, by using the MDS code, one can form $Q^{N-1}$ $N$-tuples. Moreover, we use the integers in an $N$-tuple to determine which element of the related constellation point will be chosen from which disjoint constellation among available disjoint constellations, $\mathcal{M}_1, \mathcal{M}_2, \ldots, \mathcal{M}_Q$ where $|\mathcal{M}_q|=|\mathcal{M}_{\hat{q}}|=M_1$, $\mathcal{M}_q \cap \mathcal{M}_{\hat{q}} = \emptyset $, $q,\hat{q}\in \big\{1, \ldots, Q\big\}$, and $q \neq \hat{q}$. Thus, the $N$-tuple $(I_1, I_2, \ldots, I_N)$ relates to the following set of disjoint constellations $\big\{\mathcal{M}_{I_1}, \ldots, \mathcal{M}_{I_N}\big\}$. This process produces $Q^{N-1}$ constellation points as we have $Q^{N-1}$ $N$-tuples. By using a fixed-length bit-to-symbol mapping scheme, one can map $\floor{\log_2 Q^{N-1}}$ bits to these constellation points. After determining the set of disjoint constellations, each element of the constellation points is drawn from regarding disjoint constellation. Since the size of each disjoint constellation is $M_1$ and a constellation point consists of $N$ elements, one can form $M_1^N$ different constellation points. Hence, the overall codebook size is $M=Q^{N-1}M_1^N$. Moreover, the SE per codeword element is given by    
\begin{align}
    \eta=\frac{\floor{\log_2 Q^{N-1}}+N\log_2 M_1}{N}.
\end{align}

On the other hand, one can use different approaches to determine the disjoint constellations. As in \cite{Wen2017}, one can carve the disjoint constellations from well-known PSK and QAM constellations. The MDS scheme that uses PSK constellations to form the disjoint constellations is called MDS with PSK (MDS-PSK). In this case, the first disjoint constellation is set to a regular $M_1$-PSK constellation and the remaining $Q-1$ constellations are obtained by rotating the initial PSK constellation by $2\pi\rho/(M_1Q)$ to maximize the angular difference between the constellations points \cite{Wen2017,Yarkin2021tcom}. In MDS with QAM (MDS-QAM), on the other hand, the well-known set partitioning technique is used to carve $Q$ disjoint $M_1$-ary constellations from a $(QM_1)$-ary QAM constellation.

As in IM, we can employ MDS modulation on the in-phase and quadrature parts of a codeword element separately. Assuming we have two $N$-tuples of the same MDS code for these components and each component is chosen from one of $Q$ one dimensional $M_1$-ary disjoint PAM constellations, the SE is doubled compared to the former MDS techniques,\footnote{Note that the number of disjoint constellations and size of these constellations do not have to be equal for the in-phase and quadrature parts. Here, we make such an assumption for brevity and simplicity.} i.e.,
\begin{align}
    \eta=\frac{2\floor{\log_2 Q^{N-1}}+2N\log_2 M_1}{N}.
\end{align}
To form the disjoint constellations in this technique, we follow useful guidelines in \cite{Wen2017}. Thus, we obtain $Q$ $M_1$-ary disjoint constellations by set partitioning a $(QM_1)$-ary PAM constellation. We call the resulting modulation technique MDS with in-phase and quadrature modulation (MDS-IQM). It is important to note that one can pick $M_1=1$ and do not embed information into the conventional modulation symbols for the MDS modulation techniques. In this case, each disjoint constellation has only one element. The resulting codebook consists of constellation points whose minimum Hamming distance is two \cite{Yarkin2021tcom}. In the rest of the paper, we only consider the MDS-IQM technique for ease of representation. It is also because the MDS-IQM technique is capable of achieving better error performance than the other MDS techniques \cite{Yarkin2021tcom}.  

\subsection{Receiver}
Assuming the $m$th constellation point is selected for transmission, the received signal vector at the receiver can be given by
\begin{align}
    \textbf{y}=\textbf{S}_m\textbf{h}+\textbf{w}
\end{align}
where $\textbf{S}_m=\text{diag}(\textbf{s}_m)$ and $m \in \big\{1, 2, \ldots, M\big\}$. Here, $\textbf{h}$ and $\textbf{w}$ are $N \times 1$ complex channel and noise vectors, respectively. We assume their elements are distributed with $\mathcal{CN}(0,1)$ and $\mathcal{CN}(0,N_0)$ where $N_0$ is the noise power.  

Assuming perfect channel state information at the receiver, the optimum maximum likelihood (ML) detector decides the transmitter signal vector as follows
\begin{align}\label{eq:eq11}
    \hat{\textbf{s}}=\min_{\textbf{s}\in \mathcal{S}} ||\textbf{y}-\textbf{S}\textbf{h}||^2 
\end{align}
where $\textbf{S}=\text{diag}(\textbf{s})$. 

Performing the optimum ML detector in \eqref{eq:eq11} requires excessive complexity when the size of the codebook, $M$, is high. To overcome this problem, we proposed low complexity ML (LC-ML) detectors in \cite{Yarkin2020, Yarkin2021tcom} for the MDS modulation techniques and showed that these detectors perform close to the optimum ML detector while reducing the complexity significantly.  

\section{Gray Coding} \label{sec:sec3}
It is shown in \cite{Yarkin2021tcom} that the mapping of the information bits to the tuples of the MDS code can be performed without implementing a look-up table. Hence, the transceiver complexity can be effectively reduced by implementing the mapping in \cite{Yarkin2021tcom}. We noticed that the structure of the MDS code enables Gray-like mapping when $Q$ is an integer power of two, i.e., $Q=2^l$ where $l \ge 1$ is an integer. { In this case, we have $l(N-1)$ bits to map to the $N$-tuples of the MDS code, and one can map $l$ information bits to the first $N-1$ elements of an $N$-tuple separately by using Gray coding. Thus, we map $l$ information bits to these elements in a way that information bits related to consecutive integers differs by only one-bit. In this way, we ensure a one-bit difference between the information bits related to the pairs of $(N-1)$-tuples that differs by one element and the different elements are consecutive integers. Since the pairs of disjoint constellations related to consecutive integers are closer to each other, such a mapping provides a Gray-like coding structure for the MDS modulation methods. However, this structure may not be optimum since the last integer, $Q$, rolls over to the first integer, 1, with only one-bit difference as in classic Gray coding, and the disjoint constellations related to the first and last integers are not the closest constellation pairs. On the other hand, Gray code for the tuples of the MDS code can be simply obtained by starting with an $(N-1)$-tuple of all ones and increasing or decreasing elements of such a tuple one-by-one until we have $Q^{N-1}$ different tuples.\footnote{Here, we exploit the cyclic property of Gray code where the first and the last integers are regarded as consecutive integers.} In this case, consecutive $N$-tuples will differ by only two elements and these elements will be consecutive integers.} 

\begin{table}[t!]
\centering
\caption{An example of Gray mapping for $N=3$, $Q=4$, and $l=2$. }
\label{tab:tablenew1}
\resizebox{\columnwidth}{!}{%
\begin{tabular}{|c|c|c|c|c|c|}
\hline
\multicolumn{3}{|c|}{Mapping for the First Eight $N$-Tuples}                                                                                                                                   & \multicolumn{3}{c|}{Mapping For The Remaining $N$-Tuples}                                                                                                                                     \\ \hline
\begin{tabular}[c]{@{}c@{}}The Mapping \\ in {\cite{Yarkin2021tcom}}\end{tabular} & \begin{tabular}[c]{@{}c@{}}Gray\\ Mapping\end{tabular} & \begin{tabular}[c]{@{}c@{}}$N$-Tuple\\ MDS Code\end{tabular} & \begin{tabular}[c]{@{}c@{}}The Mapping\\ in {\cite{Yarkin2021tcom}}\end{tabular} & \begin{tabular}[c]{@{}c@{}}Gray-like\\ Mapping\end{tabular} & \begin{tabular}[c]{@{}c@{}}$N$-Tuple\\ MDS Code\end{tabular} \\ \hline
$[0~0~0~0]$                                                       & $[0~0~0~0]$                                                 & $\mathcal{I}_1=(1, 1, 2)$                                    & $[1~1~0~1]$                                                       & $[1~0~0~1]$                                                & $\mathcal{I}_9=(4, 2, 2)$                                    \\ \hline
$[0~0~0~1]$                                                       & $[0~0~0~1]$                                                 & $\mathcal{I}_2=(1, 2, 1)$                                    & $[1~1~0~0]$                                                      & $[1~0~0~0]$                                                  & $\mathcal{I}_{10}=(4, 1, 3)$                                 \\ \hline
$[0~0~1~0]$                                                       & $[0~0~1~1]$                                                 & $\mathcal{I}_3=(1, 3, 4)$                                    & $[1~0~0~0]$                                                      & $[1~1~0~0]$                                                 & $\mathcal{I}_{11}=(3, 1, 4)$                                 \\ \hline
$[0~0~1~1]$                                                       & $[0~0~1~0]$                                                 & $\mathcal{I}_4=(1, 4, 3)$                                    & $[0~1~0~0]$                                                      & $[0~1~0~0]$                                                 & $\mathcal{I}_{12}=(2, 1, 1)$                                 \\ \hline
$[0~1~1~1]$                                                       & $[0~1~1~0]$                                                 & $\mathcal{I}_5=(2, 4, 2)$                                    & $[0~1~0~1]$                                                      & $[0~1~0~1]$                                                 & $\mathcal{I}_{13}=(2, 2, 4)$                                 \\ \hline
$[1~0~1~1]$                                                       & $[1~1~1~0]$                                                 & $\mathcal{I}_6=(3, 4, 1)$                                    & $[0~1~1~0]$                                                      & $[0~1~1~1]$                                                 & $\mathcal{I}_{14}=(2, 3, 3)$                                 \\ \hline
$[1~1~1~1]$                                                       & $[1~0~1~0]$                                                 & $\mathcal{I}_7=(4, 4, 4)$                                    & $[1~0~1~0]$                                                      & $[1~1~1~1]$                                                 & $\mathcal{I}_{15}=(3, 3, 2)$                                 \\ \hline
$[1~1~1~0]$                                                        & $[1~0~1~1]$                                                 & $\mathcal{I}_8=(4, 3, 1)$                                    & $[1~0~0~1]$                                                      & $[1~1~0~1]$                                                 & $\mathcal{I}_{16}=(3, 2, 3)$                                 \\ \hline
\end{tabular}
}
\end{table}

In Table \ref{tab:tablenew1}, we give an example of Gray mapping of bit strings to $N$-tuples of an MDS code when $N=3$, $Q=4$, and $l=2$. { First, we obtain the Gray code for $3$-tuples of the MDS code by increasing or decreasing the first two integers one-by-one until we have $Q^{N-1}=16$ different tuples.} Then, the mapper maps the first $l=2$ bits of the bit strings to the first element of the $3$-Tuples. The remaining  $l=2$ bits of the bit strings are mapped to the second element of the $3$-Tuples. Notice that there is only a one-bit difference between the bit strings related to consecutive integers due to the Gray mapping.  Note also that the minimum Hamming distance between the consecutive 3-tuples is two, and the bit strings related to the pairs of 3-tuples that have such a distance with consecutive integers differ only one-bit. In the table, we also provide the bit strings related to the mapping in \cite{Yarkin2021tcom} to show the difference between the mappings.

\section{Log-likelihood Ratio Calculation} \label{sec:sec4}
It is crucial to calculate log-likelihood ratio (LLR) values for each bit of the MDS modulation techniques to obtain markedly improved error performance in the presence of channel encoding. That is because soft-decision decoding based on the LLR values has a clear BER advantage against hard-decision decoding for coded techniques. Thus, in this section, we calculate the LLR values of the MDS-IQM concept. 

Since MDS-IQM employs the MDS modulation separately on in-phase and quadrature parts of the codeword elements, corresponding LLR values can be calculated for these parts independently. To better illustrate how we calculate these values, let us denote the received signal related to the $n$th codeword element by 
\begin{align}
y(n)=h(n)s_{n}+w(n)    
\end{align}
where $h(n)$, $s_n$ and $w(n)$ are $n$th elements of $\textbf{h}$, $\textbf{s}$, and $\textbf{w}$, respectively. Dividing both sides of the equation by $h(n)$, one can obtain 
\begin{align}\label{eq:eq9new}
\Tilde{y}(n)=s_{n}+\Tilde{w}(n)    
\end{align}
where $\Tilde{y}(n)=y(n)/h(n)$ and $\Tilde{w}(n)=w(n)/h(n)$. Now, let us denote $\Tilde{y}(n)$ in terms of its in-phase and quadrature components and vectors related to these components as $\Tilde{y}(n)=\Tilde{y}^I(n)+j\Tilde{y}^Q(n)$, $\Tilde{\textbf{y}}^I=[\Tilde{y}^I(1), \Tilde{y}^I(2), \ldots, \Tilde{y}^I(N)]$, and $\Tilde{\textbf{y}}^Q=[ \Tilde{y}^Q(1), \Tilde{y}^Q(2), \ldots, \Tilde{y}^Q(N)]$, respectively. Moreover, vectors related to the in-phase and quadrature components of a signal vector $\textbf{s} \in \mathcal{S}$ can be denoted in a similar way as $\textbf{s}^I=[s_1^I, s_2^I, \ldots, s_N^I]$ and $\textbf{s}^Q=[s_1^Q, s_2^Q, \ldots, s_N^Q]$, respectively, where $s_n=s_n^I+js_n^Q$. Following this notation, LLR of the $\delta$th bit, $b_{\delta}^{I}$, related to the in-phase parts of the transmitted signal can be calculated as
\begin{align}\label{eq:eq10new}
    L(b_{\delta}^{I})=\ln \frac{p(b_{\delta}^{I}=0|\Tilde{\textbf{y}}^I)}{p(b_{\delta}^{I}=1|\Tilde{\textbf{y}}^I)} \propto \ln \frac{\sum_{\textbf{s}^I\in\mathcal{S}_{\delta,0}} p(\Tilde{\textbf{y}}^I|\textbf{s}^I)}{\sum_{\textbf{s}^I\in\mathcal{S}_{\delta,1}} p(\Tilde{\textbf{y}}^I|\textbf{s}^I)}
\end{align}
where $\mathcal{S}_{\delta,0}$ and $\mathcal{S}_{\delta,1}$  are signal spaces that consist of signal vectors regarding binary bit sequences with $b_{\delta}^{I}=0$ and $b_{\delta}^{I}=1$, respectively. The proportionality at the right hand side of \eqref{eq:eq10new} is a result of the assumption that the signal vectors are equally likely. Due to the independence among the elements of the received signal vector, the likelihood probability $p(\Tilde{\textbf{y}}^I|\textbf{s}^I)$ can be obtained by
\begin{align}
   p(\Tilde{\textbf{y}}^I|\textbf{s}^I)=p(\Tilde{y}^I(1)|s_1^I)p(\Tilde{y}^I(2)|s_2^I) \ldots p(\Tilde{y}^I(N)|s_N^I) 
\end{align}
where
\begin{align}\label{eq:eq12new}
    p(\Tilde{y}^I(n)|s_n^I) = \frac{|h(n)|}{\pi \sqrt{N_0}}\exp\bigg(-\frac{|h(n)|^2(\Tilde{y}^I(n)-s^I_n)^2}{N_0}\bigg).
\end{align}

It is important to note that LLR values of the bits related to quadrature components of the transmitted signals can be calculated in the same way as \eqref{eq:eq10new}. However, \eqref{eq:eq10new} and \eqref{eq:eq12new} require $Q^{2(N-1)}M_1^{2N}$ squared Euclidean distance calculations per bit. Hence, the receiver complexity becomes dramatically high when $Q$, $N$, and $M_1$ are high. To solve this issue and reduce the receiver complexity, one can simply calculate the LLR values in an element-wise manner for the MDS modulation technique. This is only possible because each group of $\log_2 Q$ bits in $(\log_2 Q^{N-1})$-length bit sequence related to the MDS tuples are mapped to the first $N-1$ elements of an $N$-tuple code, separately. For example, let us consider the mapping example in Table \ref{tab:tablenew1} where we have $N=3$, $Q=4$, and $l=2$. If we have a look at the bit sequence $[0~0~0~1]$,  we see that the first two ($\log_2 4=2$) bits, i.e., '00', are used to determine the first element of regarding $N$-tuple as 1, whereas the second two bits, '01', are used to determine the second element as 2. The last element is $I_3=1$ since $(1+2+I_3) \mod 4 = 0$ where $I_3=1$. Hence, we divide $(N-1)\log_2 Q$ bits related to the MDS code into groups of $\log_2 Q$ bits and calculate the LLR value of the $\phi$th bit in the $\Lambda$th $(\log_2 Q)$-length bit sequence related to the MDS code as
\begin{align}\label{eq:eq13new}
    L(b_{\phi,\Lambda}^I) = \ln \frac{p(b_{\phi,\Lambda}^I=0|\Tilde{y}^I(\Lambda))}{p(b_{\phi,\Lambda}^I=1|\Tilde{y}^I(\Lambda))} \propto \ln \frac{\sum_{s_{\Lambda}^I\in \mathcal{S}_{\phi,0}}p(\Tilde{y}^I(\Lambda)|s_{\Lambda}^I)}{\sum_{s_{\Lambda}^I\in \mathcal{S}_{\phi,1}}p(\Tilde{y}^I(\Lambda)|s_{\Lambda}^I)}
\end{align}
where $\phi \in \big\{1, \ldots, \log_2 Q \big\}$, $\Lambda \in \big\{1, \ldots, N-1 \big\}$, and $\mathcal{S}_{\phi, 0}$ is the signal space that consists of one-dimensional signals with $b_{\phi,\Lambda}^I=0$. Moreover, LLR values of the $m_1$th ($m_1=1, 2, \ldots, \log_2 M_1$) bit related to $n$th ($n=1, 2, \ldots, N$) conventional PAM symbols drawn by the in-phase part of the $n$th element can be calculated as

\begin{align}\label{eq:eq14new}
   L(b_{m_1,n}^I) \!=\! \ln \frac{p(b_{m_1,n}^I\!=\!0|\Tilde{y}^I(n))}{p(b_{m_1,n}^I\!=\!1|\Tilde{y}^I(n))} \!\propto\! \ln \frac{\sum\limits_{s_{n}^I\in \mathcal{S}_{m_1,0}}p(\Tilde{y}^I(n)|s_{n}^I)}{\sum\limits_{s_{n}^I\in \mathcal{S}_{m_1,1}}p(\Tilde{y}^I(n)|s_{n}^I)}. 
\end{align}

For the information bits related to the quadrature components of the constellation points, LLR values can be calculated in the same way as in \eqref{eq:eq13new} and \eqref{eq:eq14new}. A receiver using these equations performs $QM_1$ squared Euclidean distance calculations per bit, thus such a receiver exhibit much lower complexity than that of a receiver using \eqref{eq:eq10new} and \eqref{eq:eq12new}. { However, as will be shown in Fig. \ref{fig:fig5}, a receiver performing only \eqref{eq:eq13new}  and \eqref{eq:eq14new} in the presence of channel encoding fails to provide near-optimal error performance when $N < 4$.} To solve this problem, instead of employing the MDS coding mechanism among symbol vectors, equivalent $(N, N-1)$ single parity check (SPC) code can be employed among information bits and the resulting bit sequence can be mapped to the symbol vectors. Then, extrinsic values related to the information bits can be used to improve the decoding performance. Let us again focus on the information bits related to the in-phase parts. In an $(N, N-1)$ SPC code, the $\phi$th parity check bit, $b_{\phi,N}^I$, satisfies $b_{\phi,N}^I=b_{\phi, 1}^I \oplus b_{\phi, 2}^I \oplus \ldots \oplus b_{\phi, N-1}^I$ where $\oplus$ is addition operation under modulo-2 arithmetic. For this code, it is also clear that $b_{\phi, \Delta}^I=\bigoplus\limits_{\kappa \in \mathcal{N} \setminus \Delta} b_{\phi, \kappa}^I$ where $\mathcal{N} \coloneqq \big\{1, 2, \ldots, N\big\}$ and $\bigoplus$ is the summation operation under modulo-2 arithmetic. Hence, the extrinsic value related to $b_{\phi, \Delta}^I$ can be calculated as \cite{Hagenauer96}
\begin{align}
    L_e(b_{\phi, \Delta}^I)&=L\bigg(\bigoplus\limits_{\kappa \in \mathcal{N} \setminus \Delta} b_{\phi, \kappa}^I\bigg)\\ \nonumber &= 2 \arctanh \bigg( {\prod_{\kappa \in \mathcal{N} \setminus \Delta}} \tanh{(L(b_{\phi, \kappa}^I)/2)} \bigg) \\ \nonumber & \approx \bigg( {\prod_{\kappa \in \mathcal{N} \setminus \Delta}} \text{sign}{(L(b_{\phi, \kappa}^I))} \bigg)\min_{\kappa \in \mathcal{N} \setminus \Delta}|L(b_{\phi, \kappa}^I)|.
\end{align}
Then, we use this extrinsic value to update the LLR of $b_{\phi, \Delta}^I$ as follows \cite{Hagenauer96}
\begin{align}\label{eq:eq16new}
    \Breve{L}(b_{\phi, \Delta}^I)= L(b_{\phi,\Lambda}^I)+ L_e(b_{\phi,\Lambda}^I).
\end{align}

\section{{ Numerical Results}}\label{sec:sec5}

In this section, we provide numerical BER results to illustrate the effectiveness of Gray coding and LLR calculations of the MDS methods. We also conduct numerical BER comparisons for the channel-encoded MDS and IM techniques.    

\begin{figure}[t!]
		\centering
		\includegraphics[width=9cm]{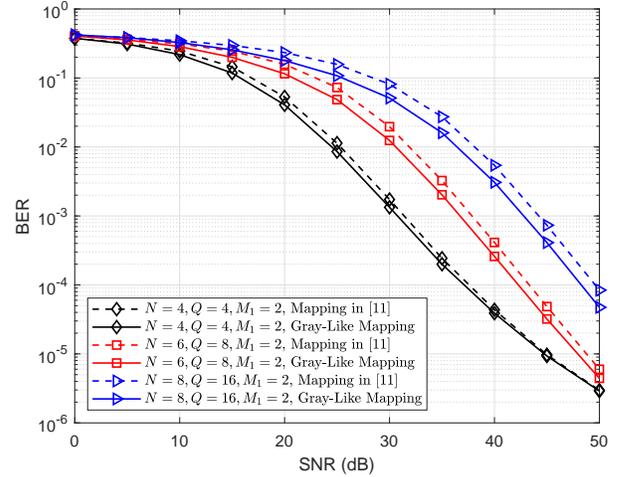}
		\caption{BER comparison of the MDS-IQM schemes employing the mapping in \cite{Yarkin2021tcom} and the Gray-like mapping.}
		\label{fig:fig1}
\end{figure}

In Fig. \ref{fig:fig1}, we compare the BER performance of the MDS-IQM schemes employing the mapping in \cite{Yarkin2021tcom} and the Gray-like mapping in this paper. Here, the BER results correspond to the LC-ML detector of the MDS-IQM schemes. As seen from the figure, the systems with Gray-like mapping (solid curves) achieve lower BER than the ones with the mapping in \cite{Yarkin2021tcom} (dashed curves). Moreover, the gap between the BER curves increases when we increase the number of $N$-tuples, i.e., $Q^{2(N-1)}$.


\begin{figure}[t!]
		\centering
		\includegraphics[width=9cm]{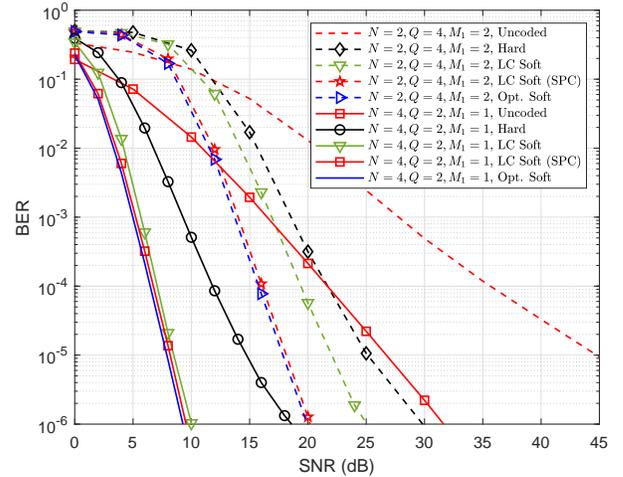}
		\caption{BER comparison of the uncoded and coded MDS-IQM schemes.}
		\label{fig:fig5}
\end{figure}

In Fig. \ref{fig:fig5}, we compare the BER performance of uncoded and coded MDS-IQM techniques.\footnote{It is important to note that we do not employ interleaving between the codeword elements of the MDS techniques related to Fig. \ref{fig:fig5} since we assume the channel fading coefficients are changing independently over each symbol interval. However, for block fading channels, one should employ interleaving to yield optimum coded performance.} For the coded schemes, we employ a rate $1/2$ convolutional code with generator matrix $G(D)=[1+D+D^2+D^3+D^6, 1+D^2+D^3+D^5+D^6]$. { Here, the BER curves of the uncoded schemes are obtained by employing the optimum ML detector, whereas those of the coded schemes with extensions ``Hard'', ``LC Soft'', and ``Opt. Soft'' are obtained by performing hard-decision decoding based on the optimum ML detector, soft-decision decoding based on the low-complexity LLR values in \eqref{eq:eq13new}-\eqref{eq:eq14new}, and based on the optimum LLR values in \eqref{eq:eq10new}, respectively. Moreover, we obtained the coded schemes with extension ``LC Soft (SPC)'' by employing the equivalent SPC code and performing low-complexity LLR values in \eqref{eq:eq13new}-\eqref{eq:eq16new}. As seen from the figure, the channel encoding improves the BER performance of the MDS-IQM techniques remarkably. The system performing only \eqref{eq:eq13new} and \eqref{eq:eq14new} suffers from significant performance loss compared to the optimum system when especially $N=2$. On the other hand, the performance of soft-decision decoder using the LLR values in \eqref{eq:eq13new}-\eqref{eq:eq16new} is very close to that of the one using optimum LLR values in \eqref{eq:eq10new}.} Such a result is critical for the MDS modulation techniques as  the complexity of calculating \eqref{eq:eq13new}-\eqref{eq:eq16new} is considerably lower than that of the optimum decoder. It also makes the MDS modulation techniques more suitable to channel encoding than the IM techniques as it is not straightforward to calculate reliable LLR values for the IM techniques.

\begin{figure}[t!]
		\centering
		\includegraphics[width=9cm]{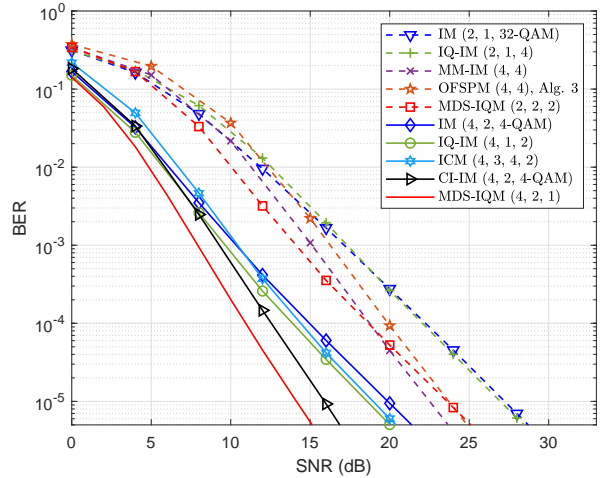}
		\caption{{ Coded BER comparison of the MDS-IQM and IM techniques.}}
		\label{fig:fig6new}
\end{figure}

{In Fig. \ref{fig:fig6new}, we compare the coded BER performance of the MDS methods with that of the IM methods.\footnote{In Fig. \ref{fig:fig6new}, ``IM $(N, K, M$-QAM)'' and ``CI-IM $(N, K, M$-QAM)'' stand for the conventional IM \cite{Basar2013} and coordinate interleaved IM (CI-IM) \cite{Basar2015} techniques, respectively, which activate $K$ codeword elements out of $N$ elements and employ $M$-QAM modulation on the activated elements. Moreover, ``IQ-IM $(N, K, M_1)$'' is the in-phase and quadrature IM (IQ-IM) technique in \cite{Fan2015} that has $K$ nonzero in-phase and quadrature components out of $N$ elements and chooses each nonzero element from an $M_1$-ary PAM constellation. ``ICM $(N, K, I, M_1)$'' stands for the index and composition modulation (ICM) scheme of \cite{Yarkin2021} that activates $K$ out of $N$ codeword elements, employs $M1$-ary PSK symbols along with the compositions of an integer $I$ into $K$ parts. Moreover, ``MM-IM $(N, M_1)$'' and ``OFSPM $(N, M_1)$'' are the multi-mode IM (MM-IM) \cite{Wen2017} and ordered full set partition modulation (OFSPM) \cite{Yarkin2020set} methods, respectively, that employ $N$ disjoint $M_1$-ary QAM constellations over codeword elements. Finally, ``MDS-IQM $(N, Q, M_1)$'' is the MDS-IQM scheme that has $N$-element codewords with $Q$ disjoint $M_1$-ary PAM constellations related to the in-phase and quadrature components.} For the MDS systems, we employ the equivalent SPC code along with soft-decision decoding based on the low-complexity LLR values given in \eqref{eq:eq13new}-\eqref{eq:eq16new}. For the remaining systems, we exploit the optimum soft-decision decoding based on Eqs. (9) and (11) in \cite{Zeng2018}. Here, all systems employ a rate $1/2$ convolutional code with generator matrix $G(D)=[1+D^2,1+D+D^2]$. Moreover, we adjust the system parameters to achieve the same level of SE for the benchmark schemes, and we used solid and dashed curves to indicate the results related to the systems achieving the same level of SE. As seen from Figs. \ref{fig:fig6new}, although the MDS concepts execute a suboptimum detection with much less complexity per bit, they outperform the IM concepts and exhibit a comparable performance to that of the MM-IM and OFSPM techniques in the presence of channel encoding. }

\section{Conclusion}\label{sec:sec6}
In this paper, we showed simple ways to perform Gray coding and low-complexity LLR calculation for the MDS-coded methods. We also stressed that these implementations are important as the benchmark techniques struggle to exhibit low-complexity structure when performing Gray coding and calculating LLR values.  We further showed by our numerical BER results that the MDS modulation schemes outperform the IM and related schemes in the presence of channel encoding.   

\bibliographystyle{IEEEtran}
\bibliography{main}
\end{document}